\documentclass[runningheads]{llncs}
\usepackage{graphicx}

\usepackage[utf8]{inputenc}
\usepackage[T1]{fontenc}
\usepackage{bm}
\usepackage[normalem]{ulem}
\usepackage{epsfig}
\usepackage{graphicx}
\usepackage{caption, subcaption}
\usepackage{algorithm}
\usepackage{algorithmic}
\usepackage{amssymb,amsmath,amsbsy,amsgen,amsfonts}    
\usepackage{dcolumn}
\usepackage{mathrsfs}
\usepackage{braket}
\usepackage{latexsym}
\usepackage{enumitem}
\usepackage{array}
\usepackage{ulem}
\usepackage{color}
\usepackage{mathtools}
\usepackage{amstext}
\allowdisplaybreaks[1]
\usepackage{txfonts}
\usepackage{pifont}

\usepackage[square,sort,comma,numbers]{natbib}
\usepackage{doi}
\usepackage{hyperref}
\usepackage{tabularx}
\usepackage[justification=justified,singlelinecheck=false]{caption}
\usepackage{epstopdf} 
\usepackage{float}
\restylefloat{table}
\usepackage{graphicx}
\usepackage[dvipsnames]{xcolor}
\usepackage{url}
\usepackage{hyperref}
\usepackage[htt]{hyphenat}
\hypersetup{
  colorlinks   = true, 
  urlcolor     = blue, 
  linkcolor    = blue, 
  citecolor   = red 
}

\usepackage[symbol]{footmisc}

\usepackage{geometry}
\begin{document}
\title{Benchmarking the Operation of Quantum Heuristics and Ising Machines: Scoring Parameter Setting Strategies on Optimization Applications
}
\titlerunning{Benchmarking the Operation of Quantum Heuristics and Ising Machines}

\author{David E. Bernal Neira\inst{1,2,3} \and
Robin Brown\inst{1,4} \and
Pratik Sathe\inst{1,5} \and
Filip Wudarski\inst{1} \and
\\Marco Pavone\inst{4} \and
Eleanor G. Rieffel\inst{2} \and
Davide Venturelli\inst{1,2,\star}}
\newcommand\blfootnote[1]{%
  \begingroup
  \renewcommand\thefootnote{}\footnote{#1}%
  \addtocounter{footnote}{-1}%
  \endgroup
}

\authorrunning{Benchmarking the Operation of Quantum Heuristics and Ising Machines}

\institute{USRA Research Institute for Advanced Computer Science (RIACS), Moffett Field, CA, USA  \and
Quantum AI Laboratory (QuAIL), NASA Ames Research Center, Moffett Field, CA, USA \and
Davidson School of Chemical Engineering, Purdue University, West Lafayette, IN, USA \and
Autonomous Systems Laboratory, Stanford University, Palo Alto, CA, USA \and
Department of Physics and Astronomy, University of California at Los Angeles, Los Angeles, CA, USA\blfootnote{\\$^\star$ dventurelli@usra.edu}}
\maketitle
\begin{abstract}
We discuss guidelines for evaluating the performance of parameterized stochastic solvers for optimization problems, with particular attention to systems that employ novel hardware, such as digital quantum processors running variational algorithms, analog processors performing quantum annealing, or coherent Ising Machines.
We illustrate through an example a benchmarking procedure grounded in the statistical analysis of the expectation of a given performance metric measured in a test environment.
In particular, we discuss the necessity and cost of setting parameters that affect the algorithm's performance. The optimal value of these parameters could vary significantly between instances of the same target problem.
We present an open-source software package that facilitates the design, evaluation, and visualization of practical parameter tuning strategies for complex use of the heterogeneous components of the solver. We examine in detail an example using parallel tempering and a simulator of a photonic Coherent Ising Machine computing and display the scoring of an illustrative baseline family of parameter-setting strategies that feature an exploration-exploitation trade-off.


\end{abstract}

\section{Introduction}
We present an approach to benchmarking the performance of hybrid quantum-classical algorithms and quantum-inspired algorithms based on a characterization of parameterized stochastic optimization solvers.
Technological progress in quantum computing and engineering has led to the proliferation of generic quantum computational methods, algorithmic applications, and hardware platforms where they can be tested.
The Noisy-Intermediate-Scale-Quantum (NISQ) \cite{preskill18} era has catalyzed a myriad of ideas and implementations of physics-based hardware approaches to optimization that do not benefit from superposition and entanglement but whose performance is nonetheless grounded in complex, difficult-to-simulate dynamics.
One class of such approaches is optimization solvers whose search algorithm is described by many coupled stochastic partial-differential equations.
These methods include analog computing with oscillator  \cite{albertsson2023highly}, and optically Coherent Ising Machines (CIMs)  \cite{mcmahon2016fully}.
Another approach is given by probabilistic bits, better known as \emph{p-bits}, an intermediate between the standard bits of digital electronics and the emerging qubits of quantum computing  \cite{camsari2019p,patel2022logically} and that can be physically implemented as perpendicular magnets.
In perhaps an abuse of terminology, these physics-based solvers are often named quantum-inspired systems.
Recent examples of the utilization of physics-inspired technologies include the design of 5G telecommunication networks via Coerent Ising Machines and Parallel Tempering  \cite{kim2021physics,singh2022ising}.
Another class of approaches are parametrized quantum circuits encoding variational algorithms such as the quantum approximate optimization algorithm (QAOA) or quantum annealing.
This type of quantum computation has been implemented on a variety of platforms, including ion traps  \cite{perez2020transitioning}, neutral atoms  \cite{dalyac2023exploring, andrist2023hardness} and superconducting qubits  \cite{kim2023evidence, maciejewski2023design}.

\vspace{-1em}
\begin{figure}[htp]
\centering
    \includegraphics[width=0.6\columnwidth]{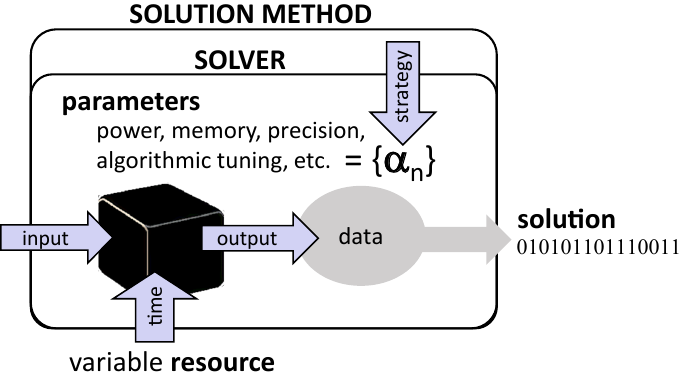}
    \caption{Abstract conceptualization of a solution method and a solver. The black box indicates the core processing optimizer (e.g., a quantum device) primarily responsible for the method performance.}
    \label{fig:solver}
\end{figure}

Empirical observations reveal that quantum and analog solvers can have an advantage over random search, producing probability distributions that potentially yield high-quality solutions  \cite{kim2021physics,king_observation_2018,coffrin2023emerging,maciejewski2023design, MohseniIsing2022}.
However, these solvers often struggle to generate samples of the global optimum and cannot guarantee its optimality, especially in the presence of noise.
Existing techniques to address this issue, such as error mitigation, primarily focus on enhancing the quality of scalar observables, such as the expectation values of functions, rather than correcting the algorithm's output (bitstrings)  \cite{kim2023evidence}.
Several other means can be employed to address this issue. 
First, improving the distribution of solution quality can be achieved through pre-processing techniques and tuning the algorithm's parameters. This is an issue for practical expected performance since good parameter settings might not be generalizable to other problem instances, success metrics, or available resources.
Moreover, the parameter-tuning strategy is resource-consuming and must be reported when discussing the solvers' expected performance.
Second, by designing solution methods that leverage the solver's capabilities to enhance the expectation value itself, the weaknesses of these methods can be mitigated by algorithmic approaches, e.g.,  \cite{brown2022copositive, dupont2023quantum}.
Assessing the performance of such methods becomes challenging as the specific sub-problems' solution only accounts for a portion of the total solution method.

As these quantum and quantum-inspired methods improve performance and capabilities, the problems they can solve become more sophisticated. 
Since the purpose of NISQ systems and quantum-inspired Ising machines is to solve 
problems, it is paramount to rigorously benchmark their performance  \cite{huckridge_benchmarking_2015}.
There is a need to develop guidelines on evaluating the performance of new computing devices in the context of their future deployment in production, i.e., guidelines for an \emph{operational} benchmarking as opposed to previous efforts that were mostly confined to research and development environments.
A full operational evaluation must include
overheads such as the cost of tuning. 
Without considering such overheads, it is easy to come to conclusions that would be misleading  \cite{AaronsonBlog}.

We propose a benchmarking framework supported by an open-source software package intended to collect statistically relevant data when running a parameterized stochastic optimization solver attempting to solve instances from a distribution of representative problems. 
This work aims to provide guidelines for presenting ``Window stickers'' - i.e., a user-friendly and self-explanatory scorecard displaying the real-world performance expectation of a self-contained method using fixed and varying resources to solve applied problems of interest.
While our considerations will mainly focus on optimization systems,  this approach can be adapted to other computational tasks (e.g., sampling, learning) and platforms (e.g., neuromorphic chips).

Algorithmic benchmarking dates back to the 70s with the work in  \cite{rice1976algorithm} on algorithm selection based on performance.
These ideas have been applied to optimization algorithms, where performance profiles  \cite{dolan2002benchmarking} are among the most popular proposals.
These diagrams show the performance of different optimization algorithms, reporting the number of problems each method can solve with respect to time.
These diagrams, although criticized for misleading conclusions when including more than two algorithms  \cite{gould2016note}, have been widely used in the literature.
Best practices have been proposed to provide the most informative benchmark analysis reported in  \cite{bartzbeielstein2020benchmarking}.
Among these best practices, an automated software tool for benchmarking ensures reproducibility guarantees, for which several tools have been proposed  \cite{bussieck2014paver,moreau2022benchopt}.

Parameter setting can be understood as an algorithm selection, where each parameterization is interpreted as a separate algorithm.
Hyper-parameter tuning and benchmarking are also relevant in other fields of computational algorithms, such as machine learning  \cite{dai2019benchmarking}, where hardware accelerators provide advantages that need to be quantified across the boundaries of different hardware implementations.
Several tools have been proposed to automate this parameter setting in that context, e.g., \texttt{Hyperopt}  \cite{bergstra2022hyperopt}.

Although there is a rich literature on algorithmic selection and parameter setting, quantum and physics-inspired optimization methods have characteristics that make their benchmarking unique and challenging  \cite{huckridge_benchmarking_2015}.
For example, new performance metrics, such as Time-to-target  \cite{king_benchmarking_nodate}, have been proposed to represent the trade-offs between solution quality and efficiency for these methods, and recently there starts to be more emphasis on measuring performance as a function of resources employed and distinguishing classes of instances  \cite{lykov2023sampling}.
Note that many quantum computer benchmarks have been mainly focused on circuit sizes that can be implemented without noise affecting their fidelity  \cite{newbenchmarking}. Additionally, in the NISQ era, quantum devices performance fluctuates over time, requiring frequent calibration. This introduces an extra layer of noise in the observed output distributions to be factored in in a benchmark.

For instance, this work aims to develop a methodology that adapts well to these quantum and physics-inspired methods for optimization, with supporting software to automate the benchmarking and parameter setting strategies and correctly account for these costs to provide practical and actionable information about solver performance. We list our contributions below:
\begin{itemize}
    \item Characterization of solution methods as instantiation of parameterized stochastic optimization solvers (see Fig. \ref{fig:solver})
    \item Proposal of benchmarking, visualizing and designing parameter setting strategies (see Fig. \ref{fig:ws_flowchart})
    \item Implementation in open-source software \texttt{Stochastic-Benchmarking}  \cite{E_Bernal_Neira_Stochastic_Benchmark_toolkit_2023}
\end{itemize}

Ultimately, the question that such a pragmatic benchmarking procedure should answer can be framed as: \emph{given well-specified resources and a new, previously unseen problem instance from a known distribution, what are the expectations for its resolution with a specific solution method?}
As discussed throughout this paper, the key to answering this question is to properly define the concepts of \emph{resources}, \emph{expectations}, \emph{solution}, and \emph{solution method}. In particular, the definition of the solution method has to address how various parameters that define the solver are set (see Fig. \ref{fig:solver}).

\section{Solution Methods and Parameterized Stochastic Solvers}

This study uses a framework focused on analyzing parameterized stochastic optimization solvers. 
Here, a ``solver" is an integrated system, where hardware (the device) and software (algorithms) work together to solve optimization problems.
Solvers have multiple parameters that can significantly affect their performance, but these effects are usually unknown beforehand.
For our analysis, the solver is seen as a sampler of random variables of an unknown distribution, a concept familiar in classical optimization as stochastic optimization methods  \cite{fouskakis2002stochastic}.
This approach is relevant for quantum heuristics and Ising machines, as they fit well within this category of optimization methods.

The raw output of such stochastic methods is a finite set, or string, of $N$ bits or binary values $\{z_i \mid z_i\in \{0,1\}\} $, obtained by a single measurement at the end of the computation
\footnote{In case of an Ising model framework, it is a spin configuration $\{\sigma_i \mid \sigma_i\in \{-1,1\} \}$, however without loss of generality, both representations are equivalent up to a linear transformation.
}, to which we associate a vector variable $\mathbf{z} = \{z_1, \ldots , z_N \}$.
The algorithm description does not specify how the distribution is updated or how the samples are obtained.

Additionally, the stochastic nature of these solvers generates a distribution of solutions, which necessitates applying postprocessing techniques to determine the required output.
A comparison between stochastic optimization algorithms and deterministic solution methods, which return the same solution to a problem every time they are executed and might even provide guarantees on the optimality of such a solution, might not be valid in general, given the heuristic nature of sampling associated with stochastic methods.
From the bitstring $\mathbf{z}$, we can define a transformed real-valued variable $X = fun(\mathbf{z})$, where $fun: \mathbb{B}^N \to \mathbb{R}$ is known as a pseudo-Boolean function  \cite{boros2002pseudo}.
\footnote{It is well known that any pseudo-Boolean function can be written uniquely as a multilinear polynomial, i.e., $fun(\mathbf{z}) = c_0 + \sum_{i=1}^N c_{1, i} z_i + \sum_{j = i +1}^N c_{2,ij} z_i z_j + \sum_{k = j + 1}^N c_{3,ijk}z_i z_j z_k + \cdots$.}
This variable, defined by a scalar function, takes the bitstring values and returns a real-valued cost or objective.
This variable $X$ can represent the solver's progress toward solving a single problem.
The solver's performance can then be assessed through variable $X$, which can subsequently be used to learn how this solver behaves across different problem instances and compared against other solvers.

Analyzing experimental results from specific cases is crucial to accurately benchmark solvers, which perform differently across various problem instances. This helps determine the solver's effectiveness for specific problem classes or families, requiring analysis over multiple instances. In studying stochastic solvers, the objective is to estimate the probability density function (PDF) of the output variable $X$ based on the samples collected during the solution process. This PDF offers empirical insight into the distribution the solver samples from.
Stochastic solvers lead to a change in the solution paradigm, where the new goal is to skew these distributions towards the desired output and sample it as efficiently as possible. 
In this perspective, deterministic solvers search over a Dirac delta distribution centered at the optimal solution; in this case, sampling becomes irrelevant, and the deterministic search becomes equivalent to finding such a distribution. 
Additionally, we focus on reporting the output for variable $X$ and the confidence level in ensuring a specific solution quality for a new, unknown problem instance within the targeted instance class.

\begin{figure}[t]
     \centering
     \includegraphics[width=0.75\columnwidth]{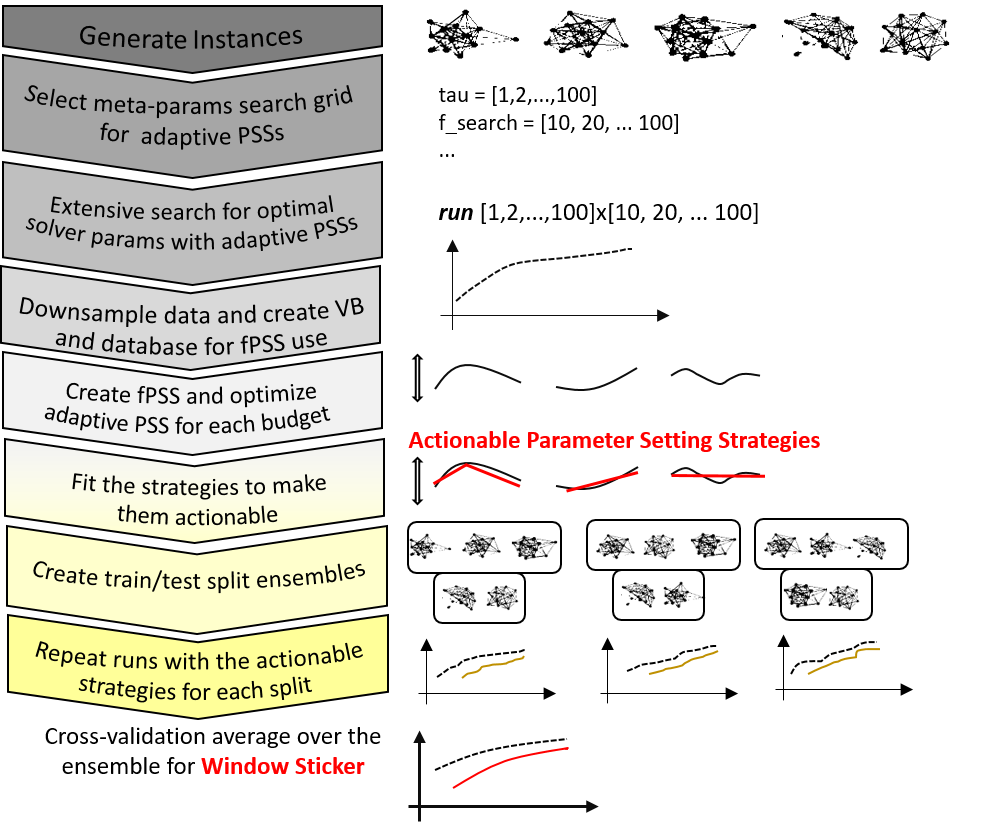}
     \caption{Flowchart with the main steps to generate the ``Window stickers'' implemented in \texttt{Stochastic-Benchmark}  \cite{E_Bernal_Neira_Stochastic_Benchmark_toolkit_2023}}
     \label{fig:ws_flowchart}
\end{figure}

\paragraph{Benchmarking Framework.} Following the concepts for a reproducible benchmark,
we present our framework as shown in Fig. \ref{fig:ws_flowchart}.
We consider an instance generation procedure, which generates a population of instances containing enough information to predict behavior over a new, unseen problem based on their solution.
This procedure is followed by selecting (meta-)parameter values for evaluating the different parameter search strategies (PSSs).
After establishing a performance metric, we set the benchmark.

The solution methods are then run by (attempting to) solve the instances to identify promising solver parameters for the figure of merit.
Recording the solution trajectory for each instance provides the information required to establish the performance profile of each method.
This information can be used to choose the best solution strategy for each value along the trajectory.
Considering this performance envelope, one could construct the virtual best (VB) performance profile.
This would be equivalent to counting with an Oracle, which can tell which solution method is best for each resource value and each instance.
This virtual best also provides a bound on the performance that can be obtained for selecting solution methods.
Moreover, by aggregating the different parameter settings that result in the best performances for each instance, one can define a fixed parameter setting strategy (fPSS).
This method is used extensively in the literature, where, e.g., the average of each parameter corresponding to the best parameter setting strategy across the different instances is computed for each resource value.
If the aggregation results in a parameter setting not initially included in the PSSs, it should be rerun for all instances to verify its performance.

One observation is that fixed parameters for the solution methods might perform suboptimally over unseen instances, as the assumption that the instance population being ``well-behaved'' or representative might fail.
A meta-parameter given to an advanced parameter tuning algorithm can address this, such as \texttt{Hyperopt}  \cite{bergstra2022hyperopt}.
These meta-parameters affect the behavior of the tuning procedure itself, as well as be used to balance the exploration and exploitation procedures of the solution method parameter tuning.
This exploration-exploitation balance can be expressed by determining which fraction of a total budget is spent looking for the best parameters and which should be spent exploiting the best-found parameter.
Moreover, during the exploration stage, each parameter setting considered could be explored for a variable amount of resources, presenting a trade-off between checking many different parameter settings or realizing the potential of each one explored after investing a larger amount of resources.

All these steps result in a trajectory of (meta-)parameters for evaluation in the solvers.
Depending on the solution methods and the family of instances, these trajectories might need to be made actionable. Namely, they might appear erratic due to a reduced number of instances or if outliers affect the different aggregations, e.g., across instances or parameter values.
Trajectories are smoothed and then rerun if they do not correspond to any evaluated PSSs to gather information about their performance.
The instance family is divided into training and testing sets to avoid overfitting the results.
The procedure for finding good PSSs is repeated over several different instance splits, and then a cross-validation scheme aggregates these results.

The resulting ``Window stickers'' consist then of parameter trajectories or plots that show the value that each parameter should follow with various resources; meta-parameter trajectories that yield the different parameter settings in adaptive PSSs, and performance profiles that show the expected merit function response to each different PSS.
These analyses can then be aggregated across different problem families to show scaling performance over a feature of the instances.

\section{The Stochastic Benchmark Framework}

\texttt{Stochastic-Benchmark} is an open-source package implementing the methodology described in the previous section   \cite{E_Bernal_Neira_Stochastic_Benchmark_toolkit_2023}.
This open-source package introduces a statistical analysis methodology for evaluating and comparing the performance of (potentially quantum and quantum-inspired) optimization solvers. 
By incorporating visual presentation techniques and robust statistical analysis, \texttt{Stochastic-Benchmark} provides researchers with a comprehensive framework to assess solver performance and facilitate informed decision-making on design and production readiness in the field of quantum and quantum-inspired optimization.
The \texttt{Stochastic-Benchmark} package holds particular relevance for analyzing quantum-inspired methodologies, which often produce a large set of solutions as outputs.
The analysis framework addresses these issues by providing a general performance comparison and parameter-setting strategy evaluation platform.

To practically implement the methodology illustrated in Fig.~\ref{fig:ws_flowchart}, we provide an efficient implementation of these methods.
In this section, we proceed to explain how the \texttt{Stochastic-Benchmark} framework operates.
Consider that the following is given: \\
- Resource to be evaluated $R = \{r_0,\dots r_f\}$. \\
- Performance metric to be considered $P$. \\
- Set of instances $I = \{i_1, \dots, i_{|I|}\}$. \\
- Set of solvers $S = \{s_1, \dots, s_{|S|}\}$. \\
- Set of pre-evaluated parameters for solver $s$, $\pmb{\alpha} = \{ \pmb{\alpha}_{1}, \dots, \pmb{\alpha}_{|\pmb{\alpha}|} \}$. \\
- Set of meta-parameters in case an adaptive PSS is to be included, \(\pmb{\theta}\).

For each solver in each parameter setting, $s(\pmb{\alpha}_n)$, a given performance profile is evaluated for each instance, $X = \text{perf}[s(\pmb{\alpha}_n), i] = P(r)$. The ordered set R of resources $r \in R$ indicates the energy, time, and memory used for each call to the solver.
Although some solvers provide the information of the performance metric as the progress of the resource, e.g., the logs provided in mixed-integer programming solvers with incumbent solutions against time, for some quantum- and physics-based methods, only the final distribution of solution is provided.
One could execute the solve for a grid of resource values, i.e., \(\forall r \in \{r_0,...,r_f\}\); however, this would be highly costly considering that access to these solvers is limited and expensive.
We implement the bootstrapping in a parallelizable manner to efficiently regenerate these profiles, using only the distribution of solutions for the largest result value \(r_f\), and compute confidence intervals for these metric predictions, which are then propagated along the data aggregations in the ``Window stickers'' framework.
By incorporating confidence intervals, \texttt{Stochastic-Benchmark} provides a robust framework for evaluating solver performance and comparing different algorithms.

The performance profiles, $\text{perf}[s(\pmb{\alpha}_n), i]$, are aggregated to compute the VB, fPPS automatically within \texttt{Stochastic-Benchmark}.
Moreover, there is an implementation to perform adaptive PSS by connecting to the hyper-parameter optimizer \texttt{Hyperopt}, and an armed bandit strategy is implemented to evaluate the balance of exploration of parameter values for solvers and exploitation of the best-found parameters.
Thus, the main idea is that with a given amount of resource budget, a fraction of those resources (\texttt{ExploreFrac}) are spent exploring the parameter space to get a sense of which parameters are suitable and then, using the knowledge obtained, spend the remaining resources running the solver with one well-informed choice of parameters.

Each of these PSS outputs a parameter strategy plot, which denotes the variation of the parameter values for different values of resources.
Actionable parameter strategy plots can be computed through callbacks in the code, which allow fitting these parameter profiles by functional forms using the Python numerical computation libraries \texttt{numpy} and \texttt{scipy}.
Finally, the software automatically partitions the instance set in the training and testing sets and repeats the benchmarking procedure for each partition, ultimately applying a cross-validation technique to tackle the parameter strategies' overfitting.

\section{Illustrative Example}
This section describes results obtained by applying the \texttt{Stochastic-Benchmark} framework on an illustrative example.
We describe the operational resources and constraints of the benchmark, the set of problem instances, the figure of merit, which information is accessible to solvers before the solution of the problems, the parameter setting strategy, and the test to assess a successful run.
We consider these to be the elements of a conscientious benchmark.

\textbf{Operational resources and constraints: Solution methods.}
We seek to minimize the energy of a class of zero-field Ising models, i.e., \(\min_{\pmb{s} \in \{\pm1\}^N} \sum_{i,j=1}^N s_i J_{ij} s_j = \min_{\pmb{s} \in \{\pm1\}^N} \pmb{s}^{'} J \pmb{s}\).
The bitstring that minimizes the problem, \( \pmb{s}^*\), and its corresponding objective or ground state energy, \(\pmb{s}^{*'} J \pmb{s}^{*}\), are desired.
For this illustrative example, we consider two solvers: parallel tempering and a chaotic amplitude control coherent Ising machine simulator.
Both methods were run on a single Ivy Bridge Node of NASA's supercomputer Pleiades, which counts with two ten-core Intel Xeon E5-2680v2 (2.8 GHz) processors per node, and 3.2 GB RAM per core, 64 GB RAM per node.
The resource considered here was the number of reads of the problem variables, also called spins given their \(\pm1\) nature, which is proportional to the time executed.

\vspace{-1em}
\begin{figure}
\captionsetup{format=plain, font=small, justification=centerlast}
 \centering
    \begin{subfigure}{0.475\textwidth}
        \includegraphics[width=\columnwidth]{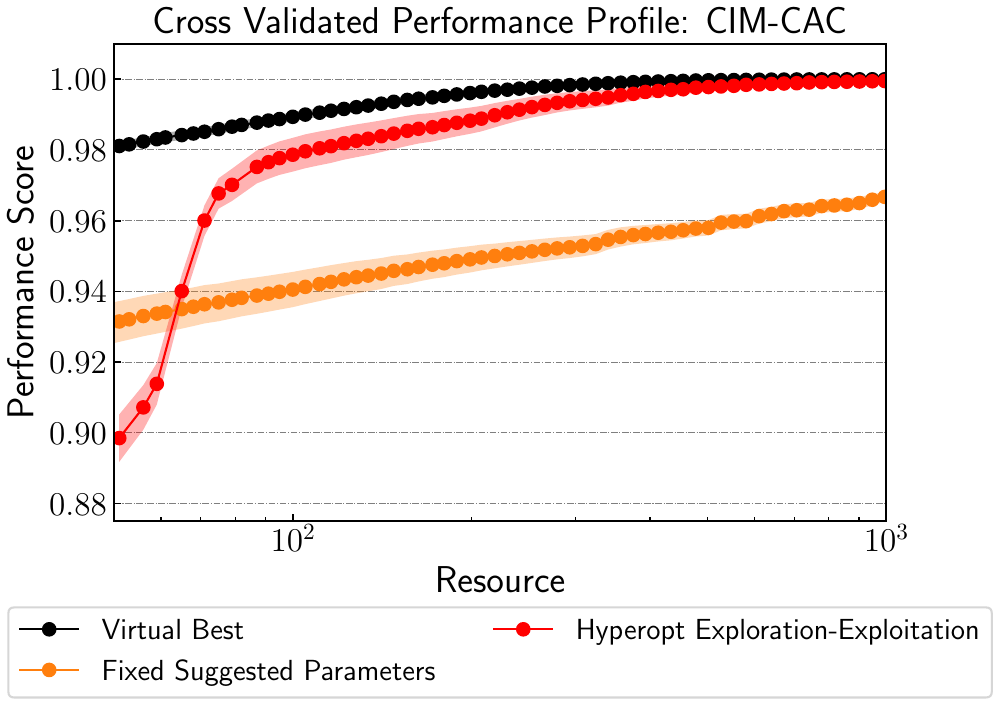}%
        \label{fig:CIM_performance}%
    \end{subfigure} \hfill
    \begin{subfigure}{0.475\textwidth}
        \includegraphics[width=\columnwidth]{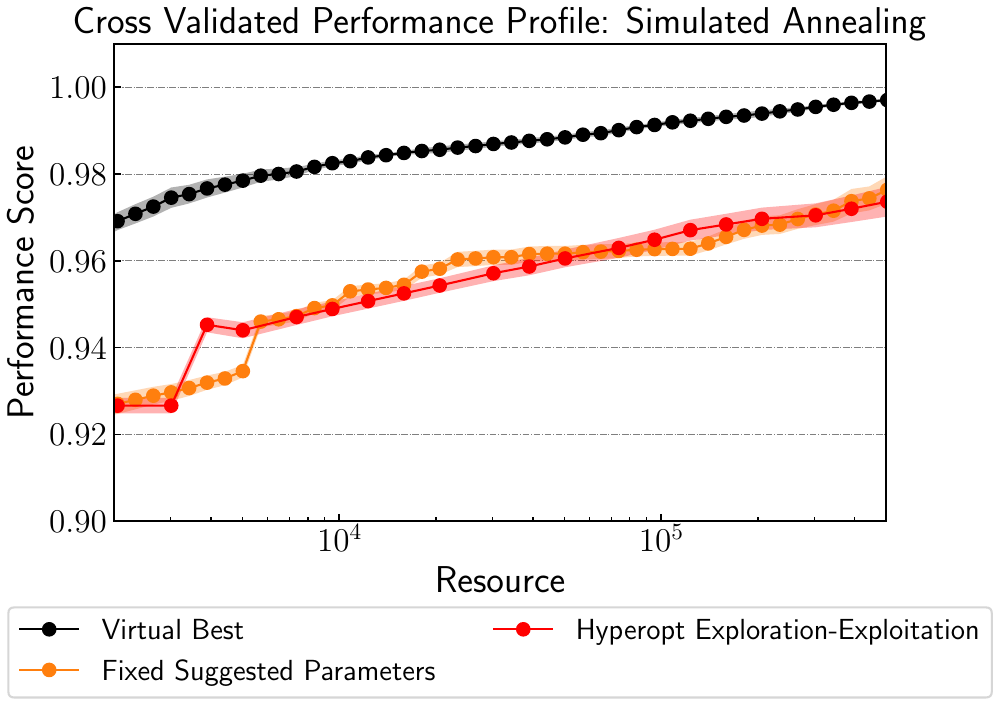}%
        \label{fig:PySA_performance}%
    \end{subfigure}
    \caption{Cross-validated performance profiles from 10 test-train splits of 50 Wishart instances with $N=50$ and $\alpha=0.5$ solved via (left) CIM-CAC  \cite{Chencimoptimizer2022} and (right) PySA  \cite{Mandra_PySA_Fast_Simulated_2023}. The profiles of the virtual best baseline, a \texttt{Hyperopt}-driven exploration-exploitation strategy, and the fixed best parameters suggested from the experiments are shown. (generated by \texttt{Stochastic-Benchmark})}
    \label{fig:performance_CIM_and_PySA}
\end{figure}
\vspace{-2em}

\textit{Solver 1: Coherent Ising Machine simulator.}
Ising machines are a class of solvers based on the dynamics of physical hardware that aims to find the minimum energy solution of the Ising model  \cite{MohseniIsing2022}.
Coherent Ising Machines (CIMs) are an example of Ising machines that exploit mixed-state density operators in a quantum oscillator network  \cite{WangCoherent2013}.
Currently, the CIM is primarily benchmarked by simulating a quantitative model of its behavior in different applications.
Although this is a widely accepted approach, no single model of the CIMs dynamics exists.
Instead, different models with varying degrees of fidelity have been constructed when modeling quantum mechanical effects.
A specific type of CIM model is called the chaotic amplitude control (CIM-CAC), which seems to provide some advantages over other types of CIM  \cite{LeleuScaling2021,reifenstein2021coherent}. Recent improvements have also emerged on the simulated model based on machine-learning insights  \cite{brown2024accelerating}.

A set of ordinary differential equations describes the CIM dynamics.
In the case of CIM-CAC, the spin variables are relaxed to continuous variables $x_i \in [-1,1]$, and auxiliary variables $e_i$ satisfy
\(\frac{d e_i}{d t} =-\xi\left(x_i^2-a\right) e_i \), 
\(\frac{d x_i}{d t} =(R-1) x_i-\mu x_i^3+\beta e_i \sum_{j=1}^N J_{i j} x_j \), 
\( a(t) =   \alpha  +  \rho \tanh (\delta \Delta H(t)) \), and
\( \xi =   \Gamma (t-t_c) \),
where \(a(t)\) denotes the squared target oscillation amplitude, and $R$ the pump schedule parameter.
After integrating these differential equations, the values of the variables $x_i$ are projected into the \(\pm1\) domain.

This solver considers four parameters, $\alpha$, $\beta$, $\Gamma$, and $R$, and the resources are given by the number of shots that account for the integration of the differential system in the time domain, simulating the execution in the hardware of the CIM.
We use a Python-based simulation library \texttt{CIM-optimizer}  \cite{Chencimoptimizer2022} to simulate CIM-CAC.

\textit{Solver 2: Parallel Tempering.}
Replica exchange MCMC sampling  \cite{HukushimaExchange1996}, which is also known as parallel tempering, is a state-of-the-art heuristic to solve Ising-like optimization problems.
Parallel tempering aims to overcome the issues faced by simulated annealing  \cite{KirkpatrickOptimization1983} by initializing multiple `replicas' at different temperatures.
The replicas undergo some Metropolis-Hastings updates, followed by a temperature swap between two replicas.
Here, we briefly describe the solver and the parameters determining the solver's performance and refer the reader to   \cite{ZhuEfficient2015a, Mandradeceptive2018} for more details.

In parallel tempering, several replicas $n_R$ are initiated at temperatures ranging between user-determined $T_{\text{min}}$ and $T_{\text{max}}$ that can be encoded in terms of two probabilities $p_{\text{cold}}$ and $p_{\text{hot}}$, that control how likely a spin flip occurs in a Metropolis update at the $T_{\text{min}}$ and $T_{\text{max}}$ respectively.
$p_{\text{cold}}$ quantifies the probability of the least likely spin flip at the lower temperature, and $p_{\text{hot}}$ denotes the likelihood of the most likely spin flip at the higher temperature.
Both these probabilities depend on the $J$ matrix values and can be approximated as 
\(p_{\text{cold}} = N_{\text{min-gap}} \exp \left( -\frac{-\Delta E_{\text{cold}}}{T_{\text{min}}}\right) \)
and 
\(p_{\text{hot}} = \exp \left( -\frac{\Delta E_{\text{hot}}}{T_{\text{max}}}\right)\)
, where 
\(\Delta E_{i} ^{\text{cold}} = 2 \min_{j | J_{ij}\neq 0}|J_{ij}| \),
\(\Delta E_{\text{cold}} = \min_{i} \Delta E_{i} ^{\text{cold}}\),
\(N_{\text{min-gap}} \)
is the count of \(E_{i}^{\text{cold}}\) being equal to \(\Delta E_{\text{cold}}\), and
\(\Delta E_{\text{hot}} = 2 \max_i \sum_{j} |J_{ij}|\).

\vspace{-1em}
\begin{figure}
\captionsetup{format=plain, font=small, justification=centerlast}
 \centering
    \begin{subfigure}{0.475\textwidth}
        \includegraphics[width=\columnwidth]{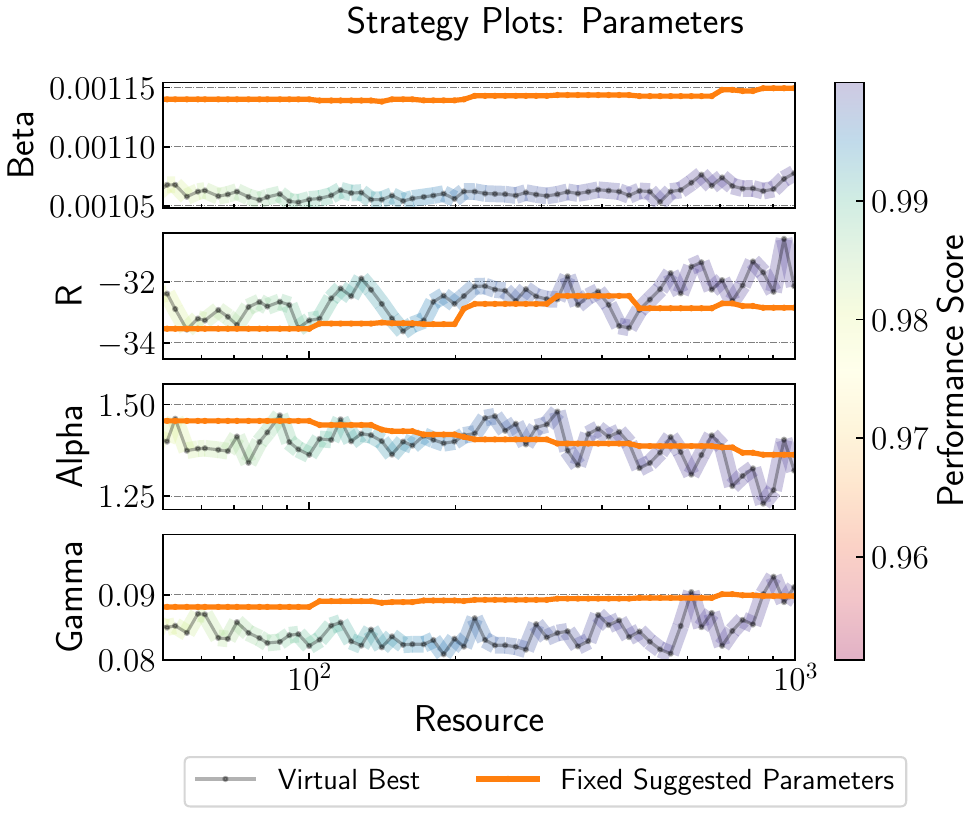}%
        \label{fig:CIM_params}%
    \end{subfigure} \hfill
    \begin{subfigure}{0.475\textwidth}
        \includegraphics[width=\columnwidth]{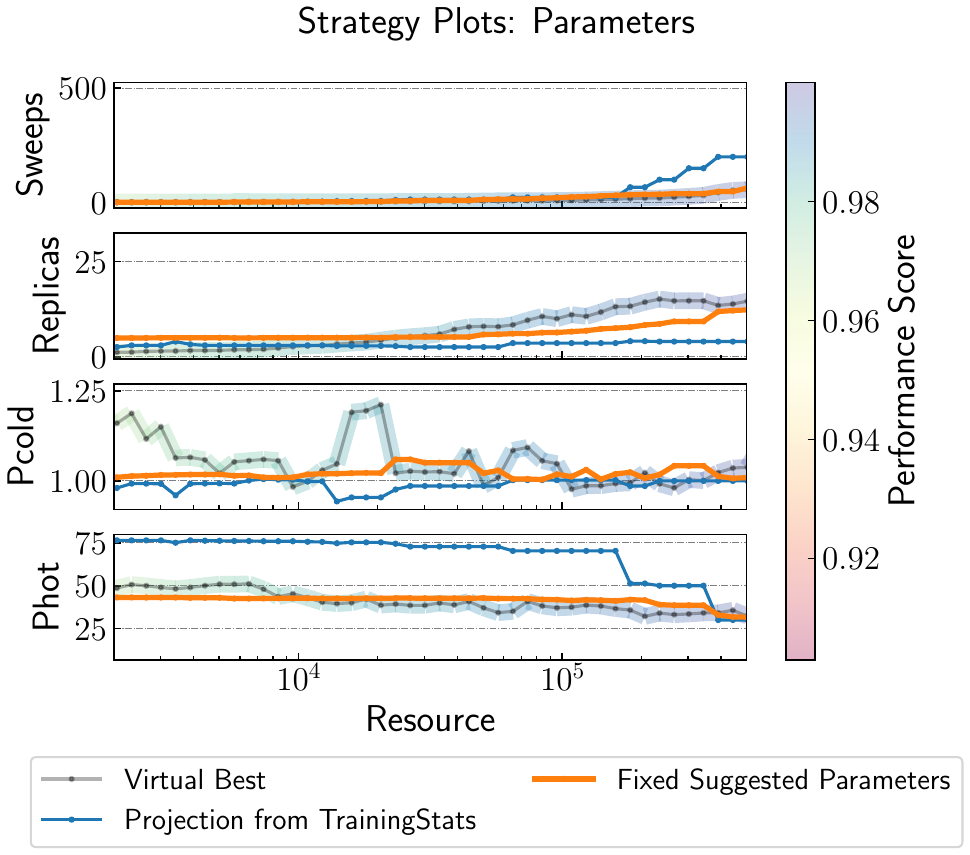}%
        \label{fig:PySA_params}%
    \end{subfigure}
    \caption{Parameter strategy plots applied to (left) CIM-CAC \cite{Chencimoptimizer2022} and (right) PySA \cite{Mandra_PySA_Fast_Simulated_2023}. Same instances and legends as Fig. \ref{fig:performance_CIM_and_PySA}}. (generated by \texttt{Stochastic-Benchmark})
    \label{fig:params}
\end{figure}

\begin{figure}
\captionsetup{format=plain, font=small, justification=centerlast}
 \centering
    \begin{subfigure}{0.475\textwidth}
        \includegraphics[width=\columnwidth]{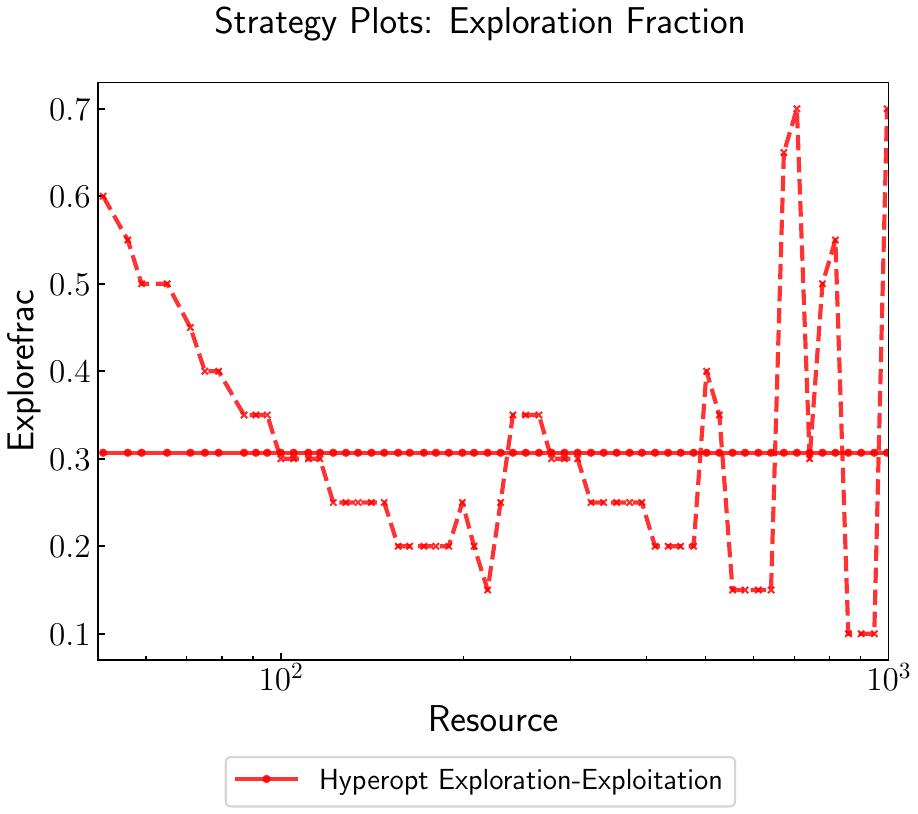}%
        \label{fig:CIM_meta_params}%
    \end{subfigure} \hfill
    \begin{subfigure}{0.475\textwidth}
        \includegraphics[width=\columnwidth]{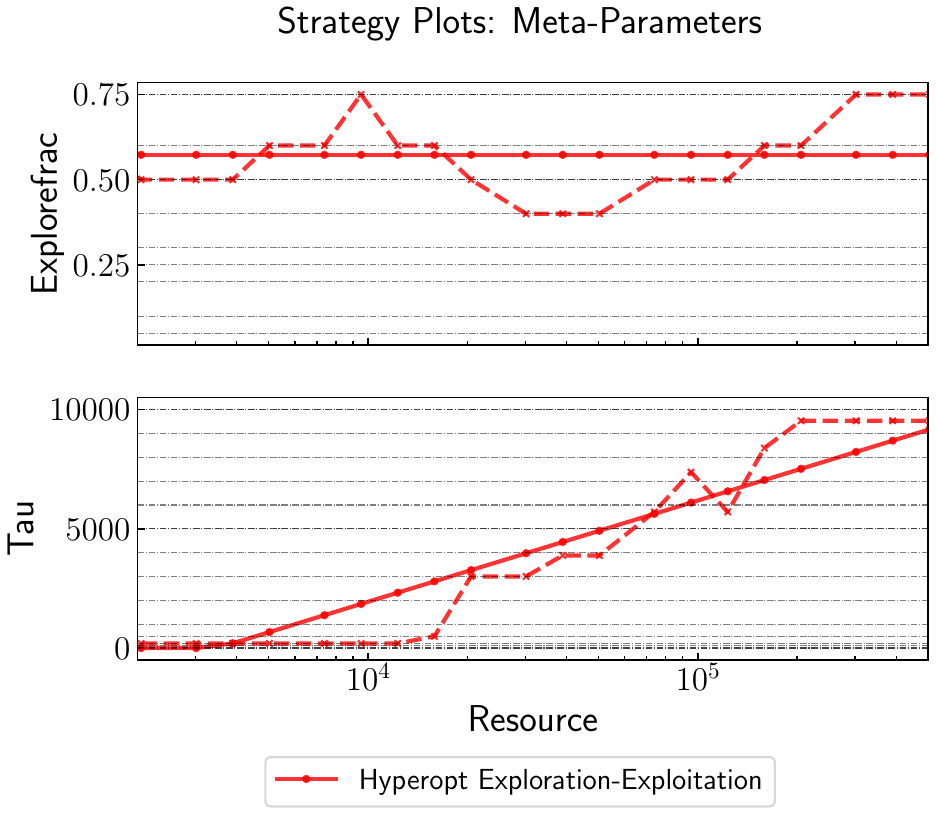}%
        \label{fig:PySA_meta-params}%
    \end{subfigure} 
    \caption{Meta-parameter strategy plots for exploration-exploitation strategy, applied to (left) the CIM-CAC \cite{Chencimoptimizer2022} and (right) PySA \cite{Mandra_PySA_Fast_Simulated_2023}. For CIM-CAC, the meta-parameter \(\tau = 1\) for all resources probed. The dashed line represents those meta-parameters with best-found performance, and the continuous line represents the actionable implementation. Same instances and legends as Fig. \ref{fig:performance_CIM_and_PySA}.(generated by \texttt{Stochastic-Benchmark}) }
    \label{fig:meta-params}
\end{figure}
\vspace{-2em}

In addition to $n_R$, the execution time is affected by another parameter, the number of sweeps $s$, which denotes the number of Metropolis updates to be implemented in the algorithm.
Thus, the solver takes four parameters, $n_R$, $s$, $p_{\text{cold}}$ and $p_{\text{hot}}$, and the resources are given by $n_R*s*$ shots, accounting for a serial execution of the replicas.
We benchmark the Python-based implementation of parallel tempering  \texttt{PySA}  \cite{Mandra_PySA_Fast_Simulated_2023}.

\textbf{Choice of Problems for Benchmarking: Wishart Instances.}
The values of $J$ are selected from the Wishart ensemble  \cite{HamzeWishart2020} to generate problem instances with planted solutions.
In particular, they correspond to the solution of the nullspace of a system of linear equations, i.e., \(W \pmb{s}^* = 0\) where \(W \in \mathbb{R}^{rows \times columns}\), out of which after a perturbation with Gaussian noise, the $J$ matrix is constructed.
The difficulty of these problems is controlled by a parameter $\alpha = rows/columns$, with a non-monotonic easy$\to$hard$\to$easy profile as $0<\alpha \leq 1$ is varied, with a critical value of \(\alpha\approx 0.2\).
We choose $\alpha=0.5$ for illustrative purposes in the following unless otherwise noted.

The Python library \texttt{Chook}  \cite{PereraChook2023} was used to generate 50 instances for each size $N$.

\textbf{Figure of merit: Performance Ratio.}
We quantify the performance using a \emph{normalized performance score} defined as follows:
\begin{align*}
    \text{Performance Score} = \frac{\text{(best found solution - random solution)}}{\text{(optimal solution - random solution)}}.
\end{align*}
Thus, the score ranges from $0$, when the solver performs no better than random sampling, to $1$, when the solver obtains the optimal solution.
Considering that we know the solution a priori (since the the Wishart instances have known solutions by design), this performance score would be closely related to the optimality gap.

\textbf{Accesible prior information.}
Although the solvers did not use any particular structure of the problems when solving the Wishart instances, their developers guided us through the ranges of the parameter values discussed below for performance.
This indication was based solely on the size of the instances, and the problem type was not revealed to the developers to avoid biases in the parameter recommendation.

\textbf{Parameter setting and run strategy.}
We provide a search space for each of the parameters considered usually over a uniform distribution around nominal values provided by the developers, except for the transition probabilities in parallel tempering, which were varied in truncated normal distributions to avoid numerical errors of the solvers.
A grid for the meta-parameters for \texttt{Hyperopt}, namely \texttt{ExplorFrac} and \(\tau\) (the resource expense of every value queried during the exploration phase) and the distributions for the parameters are reported in Appendix \ref{sec:parameters}.

\textbf{Success test.}
To obtain the performance profile (the ``Window stickers''), we analyze the performance profiles for ten test-train splits, with $80\%$ of the instances chosen as training instances and the rest as testing instances. We combine the confidence intervals and the aggregated value (mean or median) for the performance across all splits to provide cross-validated results.
The results are automatically produced by the \texttt{Stochastic-Benchmark} software and are part of the examples in the repository  \cite{E_Bernal_Neira_Stochastic_Benchmark_toolkit_2023}.

\textbf{Results.}
The cross-validated performance profiles for both technologies, obtained from 10 test-train splits of 50 instances chosen from the Wishart planted ensemble corresponding to $N=50$ and $\alpha=0.50$ are shown in Fig.~\ref{fig:performance_CIM_and_PySA}.
The framework also returns the best average values of parameters and meta-parameters for each technique and have been plotted in Figs.~\ref{fig:params} and \ref{fig:meta-params}. 
These values are intended as suggestions the framework generates to obtain the best performance on unseen problem instances.
To generate these recommendations, instead of splitting the problem instances into test and train sets, all problem instances are treated as the training set.

\vspace{-1em}
\begin{figure}
\captionsetup{format=plain, font=small, justification=centerlast}
 \centering
    \begin{subfigure}{0.475\textwidth}
        \includegraphics[width=\columnwidth]{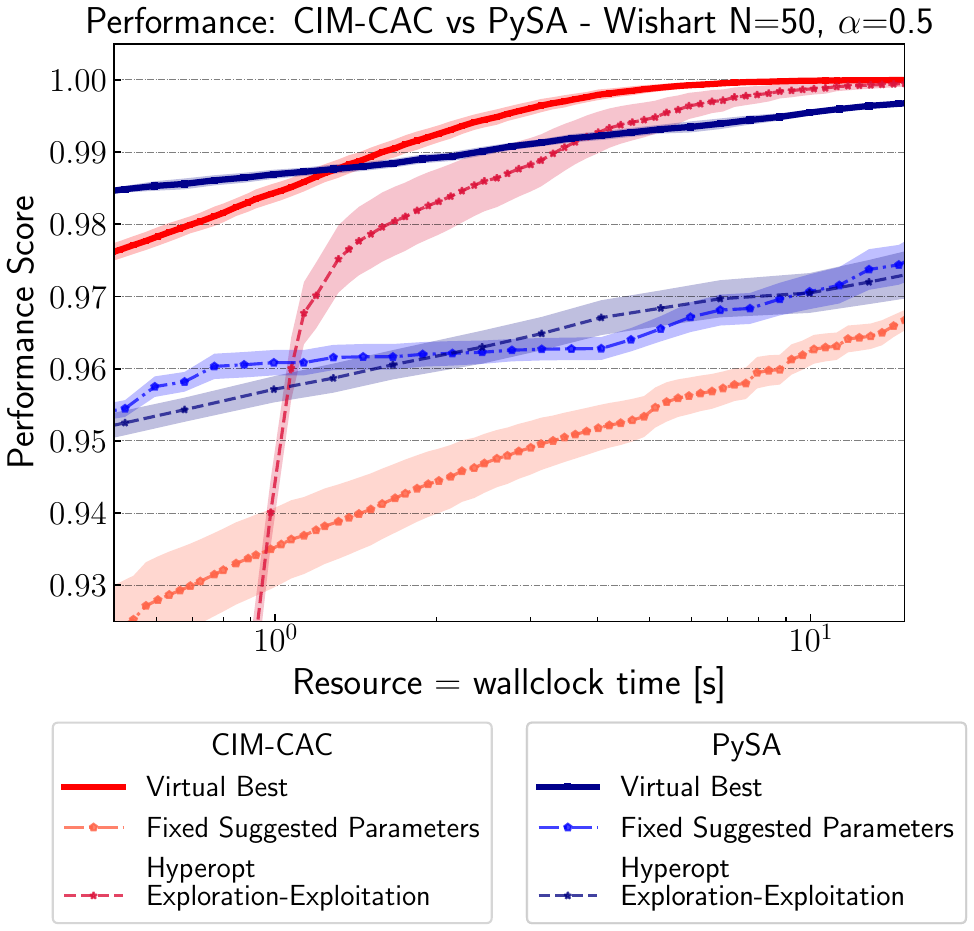}%
        \label{fig:Performance_comparison_N=50}%
    \end{subfigure} \hfill
    \begin{subfigure}{0.475\textwidth}
        \includegraphics[width=\columnwidth]{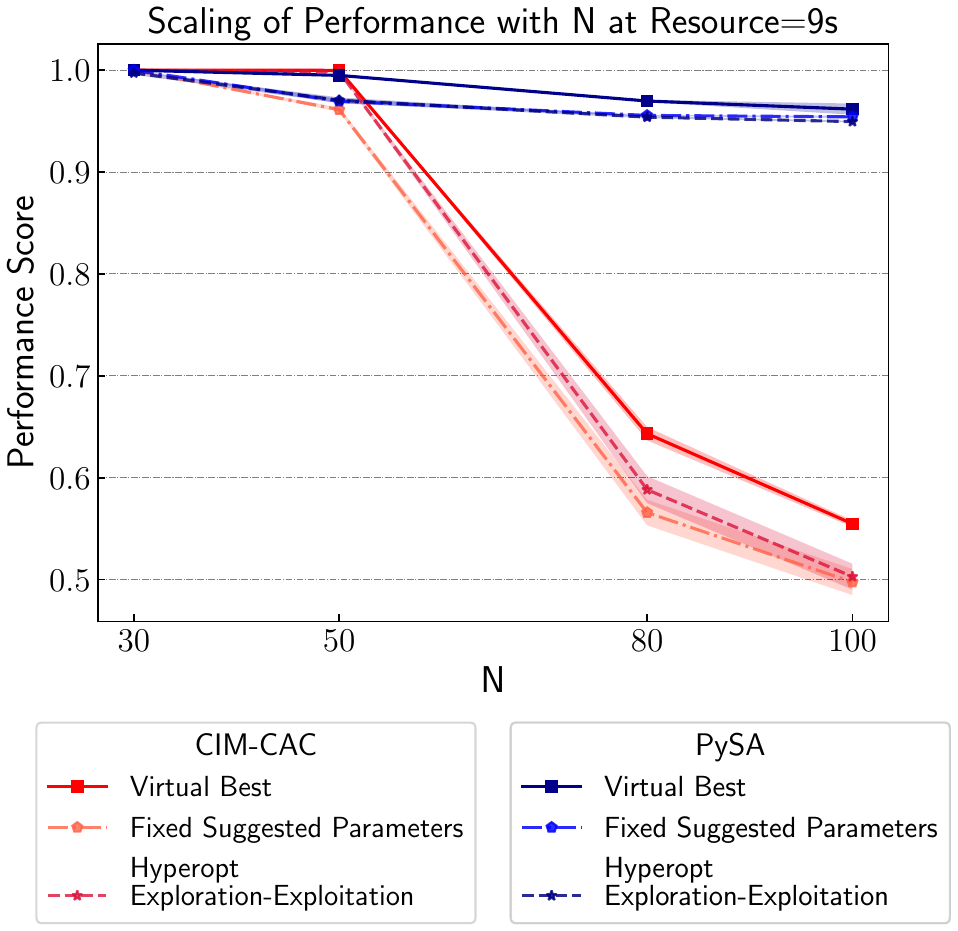}%
        \label{fig:cross_validated_scaling_both}%
    \end{subfigure} 
    \caption{(Left) Performance Comparison: The performance profiles of CIM-CAC \cite{Chencimoptimizer2022} and PySA \cite{Mandra_PySA_Fast_Simulated_2023} overlaid on the same plot, with resource chosen to be the wall clock time. (Right) Scaling of performance for both technologies with \(N = \{30, 50, 80, 100\}\). (generated by \texttt{Stochastic-Benchmark})}
    \label{fig:comparison}
\end{figure}
\vspace{-2em}

The resulting plots provide a succinct representation of large amounts of information, highlighting how to better execute these solvers when addressing new instances.
Moreover, it allows for more specialized analysis.
We include in Fig.~\ref{fig:comparison} two examples, a matching of both methods with the same resource, in this case, wall-clock time, leading to a head-to-head comparison of the methods, and an instance size scaling analysis.  By observing the N=50 results, it is apparent that in this illustrative case, our analysis allow to evaluate the benefit of using CIM-CAC with \texttt{Hyperopt} with \texttt{ExplorFrac}=0.3 and $\tau$=1 versus all other tested options, if provided a sufficient amount of resources (at least 10 seconds for this case). However, if the number of resources is not allowed to increase, it seems that PySA with a fixed PSSs is the best solver for larger problems (right plot). 
  
\section{Conclusions}\label{sec:conclusions} 
We presented an approach to benchmarking the performance of hybrid quantum-classical algorithms and physics-based algorithms based on a characterization of parameterized stochastic optimization solvers.
In addition, we introduced methods for conscientious benchmarking that provide a scheme for holistic reporting of algorithmic performance.
The analysis presented here is well fitted for stochastic optimization methods, among which we classify the quantum methods for optimization, e.g., quantum annealing and gate-based variational parametric algorithms.
The main contribution is a set of rules that characterize what an objective benchmarking procedure needs to consider, particularly with solvers spanning different hardware architectures and software that implements this for broad usage by the community.
Moreover, the methodology presented here allows for comparing different setups for a given solver, making it useful for parameter setting and tuning procedures.

\appendix

\section{Parameter values for illustrative examples}
\label{sec:parameters}

\subsection{CIM-CAC}

\textbf{Nominal values:}
\(time\_step=0.00625\),
\(R=-10.0\),
\(alpha=0.25\),
\(beta=0.0020\),
\(gamma=0.08\),
\(\delta=10\),
\(\mu=0.5\),
\(\rho=5\),
\(tau=2000\),
\(noise=0.5\),
\(T = 5000\).

\textbf{Search spaces:}
\vspace{-1em}
\begin{itemize}
    \item \(\beta \sim UNIFORM(beta, \min(beta*0.5, beta*1.5), \max(beta*0.5, beta*1.5)) \)
    \item \(R \sim UNIFORM(R, \min(R*0.1, R*10.0), max(R*0.1, R*10.0)) \)
    \item \(\Gamma \sim UNIFORM(gamma, \min(gamma*0, gamma*2.0), \max(gamma*0, gamma*2.0))\)
    \item \(\alpha \sim UNIFORM(alpha, \min(alpha*0.1, alpha*10.0), \max(alpha*0.1, alpha*10.0)) \)
    \item \(\tau = \{11, 16, 21, ...,  501\} \)
    \item \texttt{ExplorFrac} \( = \{0.05, 0.10, 0.15, ...,  1.00\} \)
\end{itemize}
\vspace{-1em}

\subsection{PySA}
\textbf{Search spaces:}
\vspace{-1em}
\begin{itemize}
    \item Sweeps:\(s \sim LogUNIFORM(10^0,10^4)\)
    \item Replicas: \(n_R \sim \text{round}(UNIFORM(1,128))\)
    \item \( p_{\text{cold}} \sim \max(logNORMAL(10^0,10^1),0.01)\)
    \item \( p_{\text{hot}} \sim \max(NORMAL(50,10),0.1)\)
    \item \(\tau = \{10, 20, 50, 100, 200, 500, 1000, 2000,5000,10000\} \)
    \item \texttt{ExplorFrac} \( = \{0.05, 0.1, 0.2, 0.3, 0.5, 0.6, 0.75\} \)
\end{itemize}
\vspace{-1em}

\small{
\section*{Acknowledgments}
We thank the NASA Quantum AI Laboratory (QuAIL) for valuable discussions, expeciallly Salvatore Mandrà, Max Wilson, and Jeffrey Marshall.
The authors thank the \texttt{CIM-Optimizer} and \texttt{PySA} developers for their advice on parameter tuning and the Pleiades supercomputer team for support in running the experiments.
This work was supported by NSF CCF (\#1918549) and NSF CNS (\#1824470) and NASA Academic Mission Services (contract NNA16BD14C – funded under SAA2-403506) and DARPA under IAA 8839 Annex 130.
R.B. acknowledges support from the NASA/USRA Feynman Quantum Academy internship program, and P.S. acknowledges support from the USRA internship program.
}

\bibliography{biblio}

\end{document}